\documentclass[conference]{IEEEtran}
\IEEEoverridecommandlockouts
\usepackage{cite}
\usepackage{amsmath,amssymb,amsfonts}
\usepackage{algorithmic}
\usepackage{graphicx}
\usepackage{textcomp}
\usepackage{xcolor}
\usepackage{fixltx2e}
\def\BibTeX{{\rm B\kern-.05em{\sc i\kern-.025em b}\kern-.08em
    T\kern-.1667em\lower.7ex\hbox{E}\kern-.125emX}}
\begin{document}

\title{Towards sustainability assessment of artificial intelligence in artistic practices \\ }

\author{\IEEEauthorblockN{Petra Jääskeläinen}
\IEEEauthorblockA{\textit{KTH Royal Institute of Technology} \\
Stockholm, Sweden \\
ppja@kth.se}
\and
\IEEEauthorblockN{Daniel Pargman}
\IEEEauthorblockA{\textit{KTH Royal Institute of Technology} \\
Stockholm, Sweden \\
pargman@kth.se}
\and
\IEEEauthorblockN{André Holzapfel}
\IEEEauthorblockA{\textit{KTH Royal Institute of Technology} \\
Stockholm, Sweden \\
holzap@kth.se}
}

\maketitle

\begin{abstract}
An increasing number of artists use Ai in their creative practices (Creative-Ai) and their works have by now become visible at prominent art venues. The research community has, on the other hand, recognized that there are sustainability concerns of using Ai technologies related to, for instance, energy consumption and the increasing size and complexity of models. These two conflicting trajectories constitute the starting point of our research. Here, we discuss insights from our currently on-going fieldwork research and outline considerations for drawing various limitations in sustainability assessment studies of Ai art. We provide ground for further, more specific sustainability assessments in the domain, as well as knowledge on the state of sustainability assessments in this domain.
\end{abstract}

\section{Introduction}

The use of Ai\footnote{The “i” is lowercase in Ai to emphasize the fact that the intelligence of current systems are quite different from human intelligence and have not yet reached a level of AGI (artificial general intelligence)} technologies has increased across different sectors of the society during the last 10 years, and immense effort and resources are currently being invested in Ai research \cite{makridakis_forthcoming_2017, bundy_preparing_2017} and development \cite{noauthor_week_nodate, noauthor_cbs_nodate}. Following the current trajectory, Ai is expected to permeate through various areas of society \cite{makridakis_forthcoming_2017, bundy_preparing_2017, minsky_future_1992}, including creative practices and creative industries \cite{noauthor_aiartistsorg_nodate, noauthor_creative_nodate, noauthor_udemy_nodate, noauthor_goldsmiths_nodate, noauthor_nvidia_nodate, noauthor_daejeon_nodate, noauthor_somerset_nodate, noauthor_serpentine_nodate, noauthor_angeles_nodate, noauthor_university_nodate, noauthor_kings_nodate, kurt_artistic_2018}. Current research into Ai and sustainability has on the other hand highlighted environmental impacts and other sustainability-related considerations that need to be taken into account concerning Ai technologies. While Ai is assumed to have many benefits, it has been suggested that we need to also think about and take stock of negative consequences of Ai both in terms of social \cite{turner_lee_detecting_2018, howard_ugly_2018} and environmental sustainability \cite{strubell_energy_2019, lacoste_quantifying_2019, dhar_carbon_2020}, ranging from algorithmic bias to the out-sized energy consumption of Ai algorithms.

Currently, there is only limited understanding of the sustainability implications of adopting Ai technologies into creative practices. The creative fields do however have direct sustainability impacts in terms of resource use and indirect impacts through their contribution to the societal and cultural fabric. The quick adaption of Ai technologies into creative practices highlights the urgency of researching the environmental sustainability of emerging Creative-Ai technologies and practices, with a variety of creative practitioners in different fields (such as poetry, music, performance, visual art).

We discuss here matters related to the environmental sustainability of Ai art. Our recent research has been focusing on exploratory virtual fieldwork to gain insights on Ai artistic practices of three Ai artists \cite{sougwen, anadol, klingemann} and have uncovered challenges in assessing the environmental impact of Ai art. We are describing here a set of factors that need to be taken into account when planning such studies. The nature of this exploratory fieldwork study has been to provide ground for further, more specific sustainability assessments in the domain, as well as to understand the state of sustainability assessments in this domain.

\section{Life-cycle of Ai artwork}

In our field studies, we have uncovered that artists' portfolios do not provide specific information on the stages, used materials, processes that lead to creating the artwork, or the maintenance and disposal of artwork. There is no clear information on which Ai models were used, the amount and kind of training data was used, or how long the training process was. This highlights the importance of establishing dialogue and research collaborations with the artists, in order to get access to this information. Following this insight, we are currently in the next phase of our research project, in which we interview artists regarding these sustainability factors of their work. The questions in our interview studies focus on used hardware, energy usage, size of models, iterations, and artists' perceptions and knowledge of the sustainability of their practice. 

In the fieldwork studies, it has also been uncovered, that there is an evident lack of scientific knowledge available on the life-cycle of Ai artwork. Life Cycle Assessment (LCA) is a common method used for analyzing the environmental impact of various technologies, but so far it has not been applied to Ai art. In arts, it has been so far used to assess the environmental impact of a media production projects \cite{persson2021} that do not involve Ai. Possibly, we can learn from these prior studies within the life-cycle assessment and apply the method for cases within Ai art. This would be particularly beneficial in contexts, where there is a potential for the art to be distributed on large scale (such as Ai-generated animations, music, literature, or deepfake media productions). As to why such assessments have not been conducted so far, we can discuss a few perspectives. One possible reason for this is the difficulty to assess artistic processes in their diversity, as the materials and stages might be very different in each project. Artists may lack awareness of the potential environmental impact of their work, and as a result, the interest to engage in such studies. In the following subsections, we discuss factors that should be taken into consideration when aiming to conduct an LCA (or other types of sustainability assessments) for Ai art. 

\subsection{Defining boundaries of the creative processes} 

In the process of looking at Ai artworks during our field study, it became apparent that many of the materials and practices do not directly contribute to artwork outcome. For instance, artists built initial small-scale prototypes to test concepts (or certain aspects of the artwork) before the construction of the final art piece. One of the artists studied was creating "re-iterations" of prior artwork that had already been exhibited, possibly using the same code and software across different artwork. This makes it particularly challenging to draw boundaries between where a piece of art begins and where it ends. If, for example, some prior artwork prototypes have contributed to the creation of a specific artwork, how should it be accounted for? To avoid such difficulties we propose that the primary aim should not be to assess individual artwork, but rather to analyze the artists' practice over a period of time. This will help us to understand if there are certain ways of practicing art that result in more environmentally friendly outcomes, and which factors in the artistic process influence these outcomes.

\subsection{Change of artwork over time}
In most cases we studied, the artists did not document their creative processes. Consequently, we know little about their work processes including the number of iterations in different phases of the creative process. This is important to understand since running an algorithm multiple times to achieve the desired outcome would result in a higher environmental impact than fewer iterations. Furthermore, if the artists altered or changed the system over time, different stages could have a varying environmental impact, making the artistic processes and the artistic practice hard to evaluate. The practice and the materials are in constant change until the point in time when the artwork is exhibited. This raises questions regarding how we should aim to assess energy usage during the creative process. What level of specificity is required for understanding the environmental impact of the Ai artwork, and can the resources used for measuring outweigh the benefits? Can a certain level of specificity be deemed sufficient? But furthermore, it shifts our focus away from the artifact, and towards the process, supporting our other findings described earlier.

\subsection{Contextual factors and supporting resources}
The artists examined here work with a variety of analog materials (\textit{e.g.} paint brushes, canvases) alongside the Ai. However, an analysis of the environmental impact that takes all these analog materials into account arguably provides lesser value for understanding the environmental impact of Ai as a creative material. However, there may for example be themes in the total use of the resources which could be of interest to study. But other, additional factors were discovered during our fieldwork: the context and supporting resources. For example, the spaces in which the performances and artwork were held can also be seen as materials of the artwork, as the artwork could not exist without these resources. The spaces used in the process of creating the artwork could also be constitutive of the environmental impact of the artwork. Other examples of supporting resources may be the travel of the artists (and the artistic team) to install and test the artifacts. Consequently, the interesting question to address is, if the spaces and resources related to Ai art differ from those encountered in other art forms in ways that affect their energy use. The contextual factors also closely relate to the matter of 'drawing boundaries' of the creative processes when aiming to assess their environmental impact.

\subsection{Temporality of display}

Regarding the display of artwork, we discovered in our field studies that in some cases the presentation/viewing of the artwork is divided into inauguration (so-called first performance) and the re-representations of the artwork at other times. The energy expenditure of this kind of re-representational viewing could be accounted for as part of a "platform" such as, for example, YouTube's environmental impact - and not the artwork itself - or it could be considered as secondary viewing of the artwork\footnote{Double accounting should obviously be avoided.}? Recently, one prominent Ai artist discussed online how the future of Ai art might see \textit{evolving} artwork, in which the artist would update the artwork online throughout the artist's life. These ideas indicate that our understanding of what it means to display and view art might change in the era of Ai art - and it also raises further questions on boundaries and what it means to "display" or "view" an artwork.

\section{Discussion and conclusion}

In our current research, we focus on studying questions related to the environmental impact of Ai in artistic practices. We have described above preliminary results that have surfaced in our online fieldwork studies. These results mainly outline the challenges in assessing the environmental impact of Ai art, but also, provide valuable knowledge and considerations for further studies in the domain. We have outlined four challenges; 1) defining the boundaries of the creative processes, 2) change in the artwork over time, 3) contextual factors and supporting resources, and 4) temporality of the display. Our next steps involve pursuing further steps in assessing the environmental impact of Ai art, understanding the artistic processes in further depth through interviews, as well as continuing to map out the knowledge landscape that could provide insights (more widely) on the sustainability of Ai art.

\bibliographystyle{IEEEtran}
\bibliography{sample-base}

\end{document}